\documentclass{egpubl}
\usepackage{sca2026}
 
\SpecialIssuePaper         

\CGFccby

\usepackage[T1]{fontenc}
\usepackage{dfadobe}  

\biberVersion
\BibtexOrBiblatex
\usepackage[backend=biber,bibstyle=EG,citestyle=alphabetic,backref=true]{biblatex} 
\electronicVersion
\PrintedOrElectronic

\ifpdf \usepackage[pdftex]{graphicx} \pdfcompresslevel=9
\else \usepackage[dvips]{graphicx} \fi

\usepackage{egweblnk} 

\usepackage{hyperref}
\usepackage{xcolor}
\usepackage{float}
\usepackage[normalem]{ulem}
\title[StayStill: a large-scale 3D idle animation dataset]%
      {StayStill: a large-scale 3D idle animation dataset}

\author[E. Atxa Landa, I. Rodriguez, E. Lazkano,  \& T. Kucherenko]
{
   Eneko Atxa Landa$^1$\orcid{0009-0003-7281-4679}, 
   Igor Rodriguez$^1$\orcid{0000-0002-1432-102X},
   Elena Lazkano$^1$\orcid{0000-0002-7653-6210}, 
   and Taras Kucherenko$^2$\orcid{0000-0001-9838-8848}
   \\
   $^1$University of the Basque Country (UPV/EHU), Spain \\
   $^2$Electronic Arts, Sweden 
}

\begin{document}

\teaser{
  \includegraphics[width=0.9\linewidth]{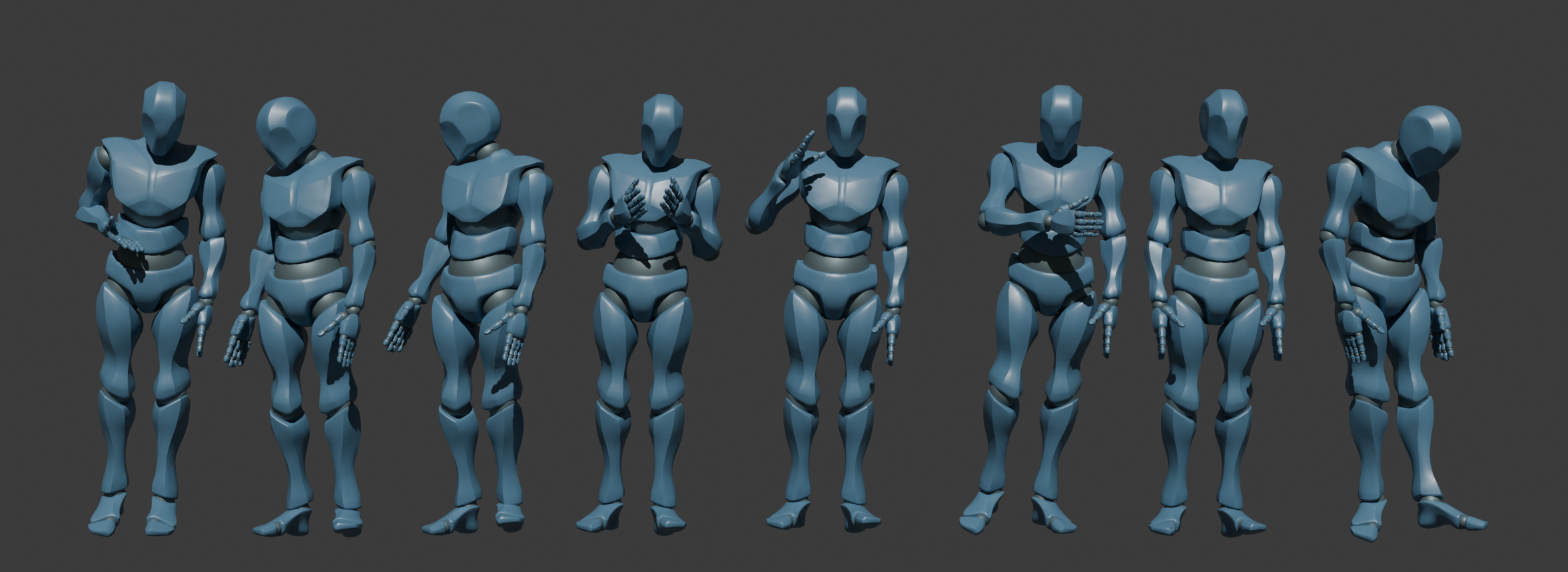}
  \centering
   \caption{Showcase of some animation clips from StayStill}
 \label{fig:teaser}
}

\maketitle
\begin{abstract}
Idle animations are essential for virtual characters, as they convey realistic behaviour during inactive states. While automatic animation generation has been widely studied, limited attention has been given to idle motion due to the absence of dedicated training datasets. We introduce StayStill, a large-scale dataset of 3D idle animations comprising diverse motion types from 50 subjects, totalling approximately 6 hours of data. We also propose an evaluation protocol for both numerical and user-based metrics as a first step towards a standardised evaluation process for future systems. To facilitate future research, we publicly release StayStill along with the evaluation code and a pre-trained baseline model that generates idle animations via transition concatenation. We believe that these contributions will enable future research on idle motion generation.


\end{abstract} 
\section{Introduction}
Idle animations are essential for the perceived believability of virtual agents. In video games and other interactive virtual environments, agents in a resting state must continue to exhibit motion in order to maintain realism. A character that suddenly stops being animated becomes obviously artificial, breaking immersion. Consequently, idle animations are a standard component of such applications and are usually created through manual animation, recorded using motion capture equipment or created using procedural animation systems.

However, the development of automatic idle motion engines and scientific research on idle motion dynamics are very limited. Other research fields related to motion generation have been extensively developed, such as human motion prediction \cite{martinez2017human, gui2018adversarial, pavllo2020modeling, hernandez2019human} or motion in-betweening \cite{harvey2020robust,qin2022motion,oreshkin2023motion,starke2023motion}. Nevertheless, these fields are usually centred around active movement scenarios such as speaking or locomotion, and the analysis and generation of idle and other low-intensity movements are often overlooked. One of the reasons for this is the lack of datasets that contain good quality idle animations, as these are essential to develop deep learning models and perform any kind of statistical analysis of motion dynamics to understand how people behave in these scenarios. 
A good quality and extensive dataset can result in the development of new research fields and enable substantial leaps in existing ones. 

For example, the field of motion in-betweening using deep learning started in 2020 with the paper \textit{Robust Motion In-betweening} \cite{harvey2020robust} by Harvey et al.
The work introduced a new research area and a first state-of-the-art technique, but perhaps more importantly, it introduced the LaFAN1 dataset. Thanks to this new public dataset that contained many types of motion (walking, running, dancing, falling, obstacle courses...) the field of motion in-betweening advanced and many new deep learning techniques were developed in the following years, such as \textit{Motion In-Betweening via Two-stage Transformers} \cite{qin2022motion}, \textit{Motion In-Betweening via deep $\Delta$-Interpolator} \cite{oreshkin2023motion} or \textit{Motion In-Betweening with Phase Manifolds} \cite{starke2023motion}. All these papers developed new techniques using LaFAN1 and furthermore, this dataset provided a standardised way to compare different state of the art techniques by providing a testing set that everyone uses.


Similarly, other foundational motion datasets, such as the Carnegie Mellon University Motion Capture Dataset \cite{cmuMocap} helped develop many research fields by providing data suitable for research. This dataset contains animations related to locomotion, sports, interaction with environment and human interaction, among others. The data became widely used in computer vision, robotics, and animation as a benchmark dataset for human motion modelling. It is one of the datasets integrated into AMASS \cite{mahmood2019amass}, a dataset with wide adoption in the animation community.

In order to develop a deep learning-based idle animation system, two elements are needed: a dataset to train the models on and an evaluation protocol to track improvements. To address these requirements, this work provides two primary contributions:

First, we introduce StayStill, an openly distributed idle animation dataset, which is, to our knowledge, the first large-scale 3-dimensional animation dataset focused on idle animations. StayStill contains around 6 hours of motion from 50 different people to ensure diversity. On the one hand, it contains 2-minute-long sequences of different participants performing natural idle animations (both without and with a phone involved), and on the other hand, what we refer to as ``idle actions'': a set of 18 different annotated typical movements related to idle motion, such as scratching different body parts, looking around, stretching or checking a watch. We provide manually annotated and cleaned motion, to improve the quality and usability of the data.

Secondly, we take the first steps towards standardised evaluations for idle animation generation by proposing an evaluation protocol, based on user-based benchmarking with 118 participants and a numerical evaluation. Motion in-betweening on idle motion is selected as the target task, baseline techniques are compared against a transformer network trained on StayStill. Widely used numerical metrics are reported on a deep learning-based state-of-the-art network and some other baselines. A user study is also conducted to make this comparison based on human perception. To support reproducibility, we openly provide the code for the numerical and user-based evaluations, the response data from the user-based evaluation, the pre-trained model and the code to generate the animations in the project page (\url{https://enekoassets.github.io/staystill.html}). 

With this paper, we aim to enable machine learning-based idle motion generation, since proposing and openly publishing a new large-scale dataset is crucial in order to develop any new research field, as it has previously been seen in motion in-betweening and motion synthesis. Moreover, we believe that having a standardised evaluation pipeline based on user studies is the next logical step to develop high quality motion generators.

The remainder of this paper is organised as follows: in Section \ref{sec:related_work} we review previous work in idle motion generation and motion in-betweening. Section \ref{sec:dataset} introduces StayStill by detailing our data collection process and recording methodology, followed by a comprehensive description of the resulting motion sequences. In Section \ref{sec:data_validation} we perform many experiments to validate the usability of the data for training deep learning networks. Finally, in Section \ref{sec:conclusions} we make a discussion about several aspects of the dataset and define the path for future work.

\section{Related work}\label{sec:related_work}
\subsection{Idle motion generation}
The scientific literature in the field of idle animation generation is currently very scarce. Moreover, these applications cannot be found publicly, and they date back to 2004. Egges et al. \cite{egges2004personalised} generated idle motion based on Principal Component Analysis, by combining change of balance with small posture variations. In \cite{egges2004example} they further developed an idle motion engine with a user interface, which generated idle motion by blending pre-recorded animations. Kocoń \cite{kocon2013idle} also developed a system that generated idle motions using kinematic chains of rigid elements applied to a human head model.

When it comes to datasets, the most notable idle data collection effort is IdlePose \cite{ravenet2021idlepose}. In this paper, the author presents a methodology to capture genuine idle motion using a deception technique. The idea of the recording process and the execution is well documented and the ethical implications are correctly addressed. However, IdlePose has two main drawbacks: firstly, there are no links in it to online repositories from which the dataset could be downloaded. Secondly, it was recorded using a single camera and the keypoints were extracted using pose estimation software. This means that the dataset is 2-dimensional, while most applications nowadays require 3-dimensional animation data.

In our previous work \cite{landa2025evaluating}, we introduced ReActIdle, a 3D idle animation dataset. However, its size is relatively limited (around 45 minutes), since the primary contribution of the paper is validating that genuine and acted idle motions are perceptually indistinguishable. The dataset was created to support this hypothesis; therefore, it is not intended for large-scale applications, such as training deep learning models, and it primarily contains general idle motion sequences. In contrast, the current paper proposes a much larger dataset, featuring more diverse idle motion classes with greater variability, which is better suited for a wider range of applications, including deep learning model training.

Other online animation repositories contain idle animations, such as Mixamo \cite{mixamo}, which provides around 15 idle animation loops. However, these repositories are not suitable to train deep learning systems due to end user license agreements, and more importantly, because they don't have a high volume of idle animations. Other commercial non-free alternatives include idle sequences, often bundled alongside other animations, such as the Unity Asset Store, the Unreal Engine marketplace, Actorcore or Synty, among others.

With the current state of public idle motion datasets, it is virtually impossible to develop any kind of idle motion synthesiser, especially if using deep learning is intended. Any of these efforts would first require recording idle motion, which can be costly and time consuming.

\subsection{Motion in-betweening}
As stated before, a common way of motion generation is motion in-betweening. In this task, given an initial context and a final target frame, a model completes the motion between the two by generating the necessary frames. This type of techniques enables professional animators to automatically generate an initial animation to visualise and work over. Motion in-betweening has been used in this paper as a base to develop an idle animation generator, by training a network to generate natural looking transitions between keyframes, and concatenating these transitions into a realistic long idle sequence.

The first deep learning–based approach to this problem was introduced by Harvey et al. \cite{harvey2020robust} in \textit{Robust Motion In-betweening}. This aforementioned foundational paper presents a recurrent adversarial network to tackle the problem, augmented with two additional components: a \textit{time-to-arrival embedding} and a \textit{scheduled target noise} vector. The authors evaluate their method on the Human3.6M \cite{ionescu2013human3} dataset, and introduce the novel LaFAN1 dataset, specialised for motion in-betweening. The impact of LaFAN1 on subsequent research in motion in-betweening has been substantial, serving as a primary motivation for the present study, which likewise aims to publicly release a dedicated dataset for idle motion generation.

Following this first work, Qin et al. \cite{qin2022motion} present a two-stage pipeline using two individually trained transformers to generate a coarse in-betweening prediction and refine it for more realistic results. In \cite{oreshkin2023motion}, Oreshkin et al. present a transformer network that refines the output of a SLERP interpolator. 
Starke et al. \cite{starke2023motion} use a mixture-of-experts model with phase variables learned by a Periodic Autoencoder. Akhoundi et al. \cite{akhoundi2025silk} perform motion in-betweening with a simple Transformer-based network. They presented strong results and found that data modelling choices are crucial to improve performance, such as increasing data quantity, the choice of pose representation and incorporating velocity input features.

The LaFAN1 dataset serves as the primary benchmark for the cited literature. By providing a specialised dataset for motion in-betweening, with standardised train-test splits, this dataset has facilitated iterative improvements and rigorous comparative analysis of emerging techniques. Inspired by the utility and accessibility of LaFAN1, we collected the StayStill dataset aiming to catalyse similar advancements within the field of idle animation synthesis.

\section{Dataset}\label{sec:dataset}
According to \cite{landa2025evaluating}, genuine and acted idle motion data is perceptually equal. In other words, when observing an idle animation, users are generally unable to determine whether the person that is performing has been asked to act as if they were idling or if they have been recorded without them noticing. This means that explicitly asking people to idle and recording them gives equally convincing results as in-the-wild capture. For this reason, the technique employed in this paper does not use a deception based technique to record the participants, instead, they were fully aware of being recorded.

\subsection{Data collection}
The data collection process has two important aspects that need to be discussed: the hardware and software setup used to obtain the motion data; and the recording methodology used for all the participants, in order to obtain a diverse and extensive dataset. 
\subsubsection{Hardware and software setup}
Recording idle motion can be challenging because of the nature of idle movements themselves. The most common approach to obtain high-quality animations is the usage of motion capture (mocap) suits and professional actors. These are instructed to act different movements while wearing a mocap suit. Then, by using cameras or different types of sensors, such as inertial measurement units (IMU), a software is able to output a 3-dimensional animation.

However, because high-end motion capture systems are expensive, markerless mocap provides a more cost-effective and accessible alternative, facilitating reproducibility. Due to the lack of access to a dedicated motion capture studio, we employed a markerless mocap setup. Specifically, we used the Freemocap \cite{matthis2022freemocap} software, an open-source tool that leverages multiple cameras and Mediapipe’s \cite{lugaresi2019mediapipe} pose estimation framework to generate 3-dimensional animation data.

\begin{figure}[htb]
    \centering
    \includegraphics[width=\linewidth]{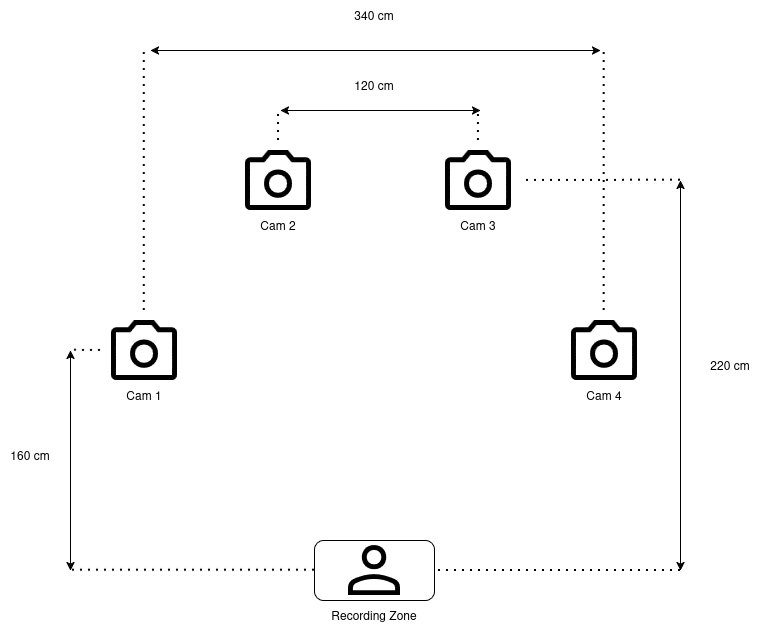}
    \caption{Diagram of the camera setup}
    \label{fig:camera_setup}
\end{figure}

The recording setting consisted of 4 GoPro Hero 11 cameras, located in a semicircle in front of the participant, directly pointing to the recording area. Figure \ref{fig:camera_setup} shows the approximate measures of the camera placement, although Freemocap is able to handle different position and rotation settings using a ChArUco board calibration. The 4 cameras record the same scene simultaneously in 1080p at 30 fps, to later detect a 2-dimensional skeleton in each video. Finally, it uses triangulation to reconstruct the 3-dimensional skeleton.

With this hardware setup, Freemocap has two requirements: on the one hand, all 4 videos need to be synchronised and have the same number of frames; this has been manually done with the help of a visual clapperboard. On the other hand, a calibration process is required to determine the intrinsic and extrinsic parameters of each camera. This calibration process has also been manually done before each recording session to ensure maximum precision.

\subsubsection{Recording methodology}
The recording process consisted of three phases, each containing a different type of animation. Before starting the recording process the entire procedure was explained and discussed with each participant, in order to ensure that each one of them understood all the parts before recording. After the explanation, the participants read and signed a consent form in which their voluntary participation was formally determined, and their right to access and remove their data was presented. The recording experiment was approved by the ethical committee of the university. Below, each one of the three phases is described in more detail.

\begin{itemize}
    \item \textbf{Phase 1 (idle actions)}: In this phase, participants were asked to act 18 actions related to idle movements. These actions were the following: looking up at the sky/the weather, looking around/the street, looking at the floor, looking and adjusting one's shoes, checking a watch, unlocking and checking the mobile phone, scratching the head, scratching an arm, scratching a leg, scratching the back, touching the face or chin, rubbing the eyes, yawning, looking to the back from the right, looking back from the left and changing the balance from left to right and from right to left. Each action was performed 2 times by each participant in order to capture more variability. There were no restrictions in this phase, so the 2 actions could be performed in a similar manner or in a completely different one. 
    \item \textbf{Phase 2 (general idle)}: In this phase, the participant was asked to act general idle motions. To better explain this, an imaginary situation was communicated to them, in which they were waiting in the street for someone or for a bus. The phase lasted for 2 minutes, in order to achieve an acceptable size for the dataset and obtain long sequences of idle motion, while still trying to avoid unnatural and repetitive movements. The only constraints were that participants could not use a mobile phone and had to remain within the recording area. They were permitted to move their legs freely but were required to remain stationary, without walking.
    \item \textbf{Phase 3 (idle with a phone)}: In this last phase, participants were again asked to act as if they were waiting in the street, but in this case they were asked to use their mobile phone in the meantime. This phase was conducted since using a phone while idling is a very standard way to idle nowadays, and this type of data could be useful for various applications. This data enables the development of more robust animation generators that incorporate phone usage behaviours. Such generators can be applied in modern video games that portray non-playable characters interacting with phones, such as \textit{The Sims} or \textit{Grand Theft Auto}. This third part was also 2 minutes long, and the only restriction was not to walk out of the recording zone.
\end{itemize}

The main reason to record the idle actions before everything else is that it is easier to perform specific actions than to respond to the more general instruction of ``act as if you were waiting". By recording these actions before everything else, participants had some time to adapt to the recording setting and they also gained a better understanding of the nature of idle motion.

\subsubsection{Data cleaning and annotation}
We acknowledge that a markerless capture system may introduce more motion artifacts than motion capture suits. Foot sliding, self-penetration and jittery motion are usually more present when using markerless systems.  In order to minimise the motion artifacts that are present in StayStill, after obtaining the raw animation files from the recordings, a post-processing phase was conducted. This phase was performed by using the \textit{bvhTools} \cite{bvhTools} library, which is an open-source python library that enables to easily perform many operations with BioVision Hierarchy (BVH) animation files, such as reading, writing, editing, visualizing or extracting statistics from them.

Firstly, the idle actions from the first recording phase were manually divided into individual animation clips, each containing an entire individual action, and annotated with their corresponding label. The general idle and the idle with a phone animations were also separated. Then, each animation piece was centred, so in the first frame the skeleton stands in the $(0, 0, 0)$ position.

In the cleaning phase, each idle action clip was manually revised, and faulty animations were removed. Clips that contained multiple simultaneous actions were also discarded: for example, if a person was asked to look at their watch, but in the meantime they scratched their head with the other hand (an entirely normal behaviour), the clip was discarded as it did not strictly contain the intended action.

Finally, each idle animation was carefully manually inspected, and faulty sections that occur because of incorrect detections of the pose estimation were located. Instead of directly removing these parts, we provide the annotation file along with the code to easily remove them, so the final user is able to make the final choice of removing the data or keeping it. Additionally, we provide a more aggressive automatic cleaning script that removes high-speed segments to yield a more refined dataset, albeit at the cost of increased data loss. Foot sliding and self-penetration are artifacts that are more difficult to detect and filter out, so they might be present in some sections of StayStill, although the manual cleaning filtered out part of the foot sliding artifacts.

\subsection{Dataset description}
\subsubsection{Motion data}
The final clean dataset contains 645.338 frames of motion at 30 fps, totalling 5 hours, 58 minutes and 31 seconds. The idle actions part consists of 275.751 frames, the cleaned general idle part is comprised of 181.846 frames and the idle with a phone part contains 187.741 frames. All the motion comes from 50 different non-actor subjects, ranging from 22 to 65 years old, 70\% male and 30\% female, to ensure varied ways of performing actions and standing still.
Table \ref{tb:motionData} (Appendix \ref{app:motionData}) shows the recorded motion types, durations and clip quantities in more detail. The quality of the data can be inspected on the video provided in the Github repository.

\subsubsection{Skeleton structure}
Freemocap provides an output skeleton of 63 bones, 40 of which correspond to the fingers. We remove the finger bones, as they may introduce noise because of the difficulty of correctly detecting fingers introducing faulty detections.

\begin{figure}[htb]
    \centering
    \includegraphics[width=\linewidth]{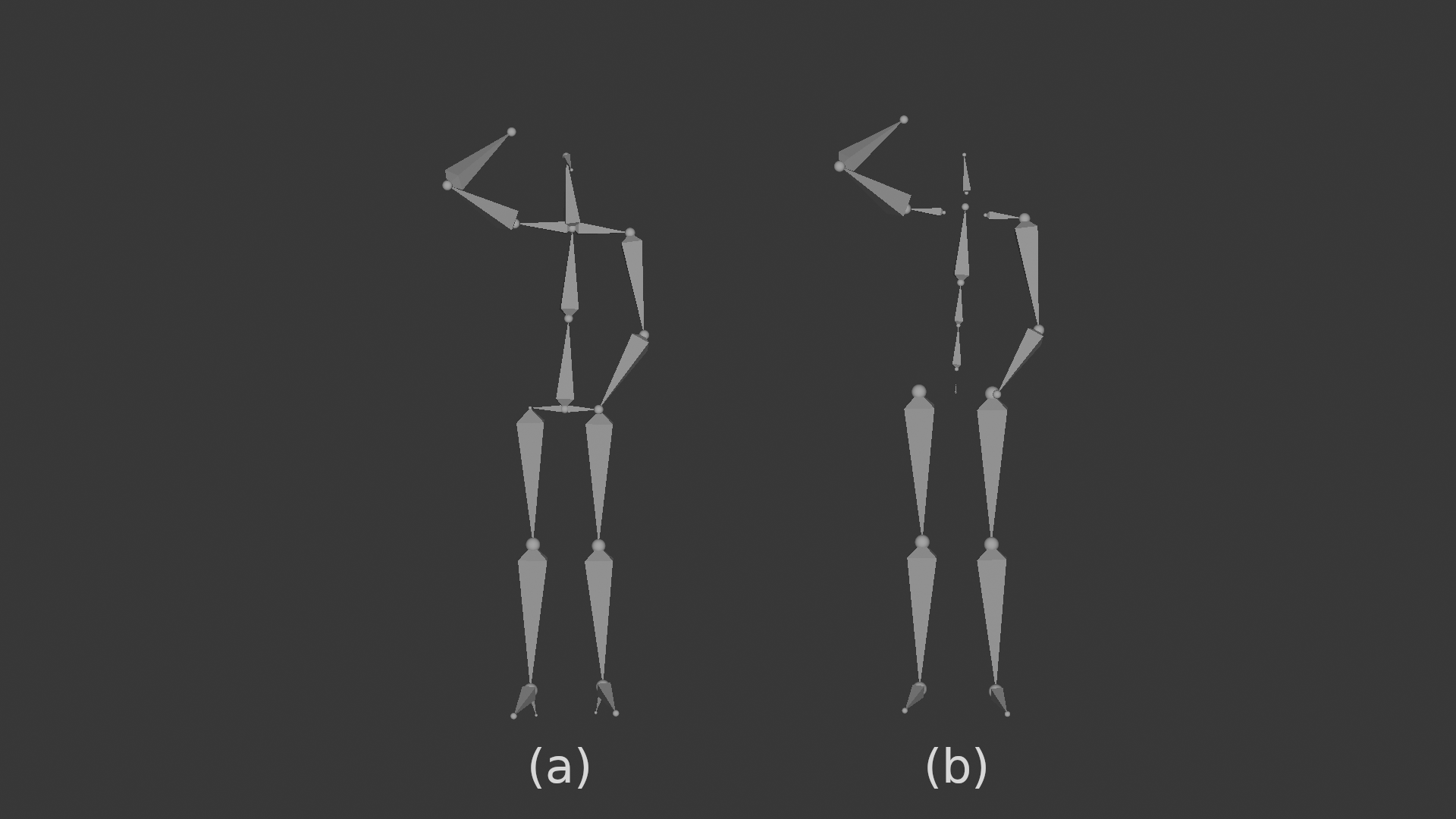}
    \caption{A frame of motion on the Freemocap skeleton without hands (a), the same motion frame on the LaFAN1 skeleton (b)}
    \label{fig:skeletons}
\end{figure}

The dataset has also been retargeted to the same skeleton format that the LaFAN1 dataset uses, which consists of 22 bones. Having the data available in this format makes combining the two datasets much easier in case it is needed and it also permits to use existing state-of-the-art techniques that use LaFAN1 directly on StayStill, without the need of changing architectures. The retargeting process has been carried out with the open-source blender-animation-retargeting plugin by Mwni \cite{blenderRetargeting}. Figure \ref{fig:skeletons} shows an example frame from the dataset on the final Freemocap skeleton (a) and on the LaFAN1 skeleton (b).

\subsubsection{File format and examples}
\begin{figure}[htb]
    \centering
    \includegraphics[width=\linewidth]{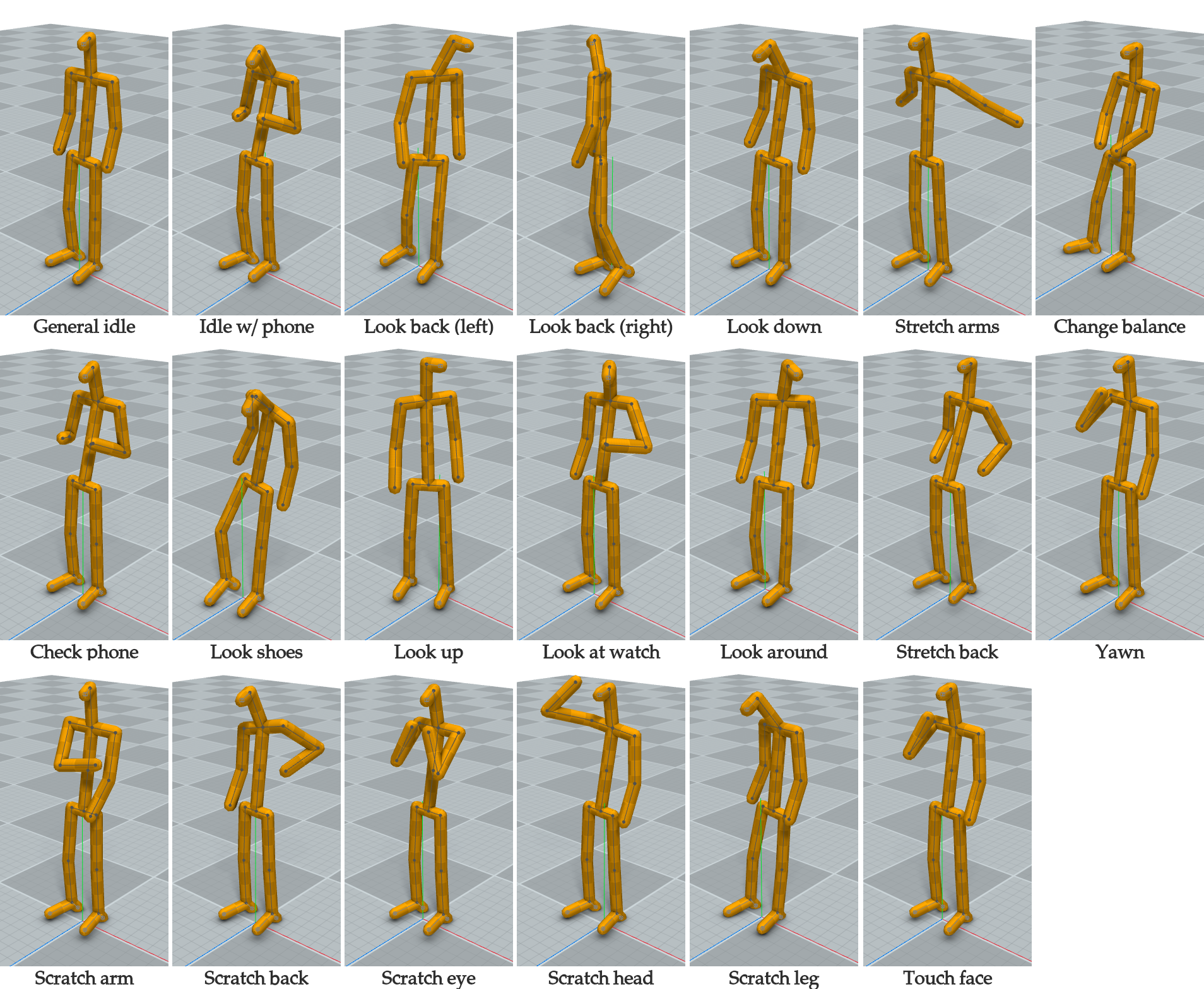}
    \caption{Example frames of different classes present in the dataset, rendered using capsules, in bvhView \cite{bvhView}}
    \label{fig:examples}
\end{figure}

All the recorded motion is provided at 30 frames per second, in the widely extended BVH format. BVH files contain a header section that defines the hierarchy and shape of a skeleton, followed by a motion section, containing the rotation values for each bone in each frame. Figure \ref{fig:examples} contains some significant example frames contained in the dataset.

\subsubsection{Standardised dataset split}\label{sec:officialSplit}
To facilitate reproducible evaluation and fair comparisons across future works, we propose a standardised train-validation-test split for StayStill. Following the practices of LaFAN 1, we use subject-wise splitting, as it better reflects the generalisation capabilities of the systems. Moreover, we propose a 70\% train, 10\% validation, 20\% test approach, to keep the same test set proportion:
\begin{itemize}
    \item Train subjects: 0 1 3 4 5 7 11 12 13 14 16 17 18 20 24 25 26 27 29 30 31 32 33 34 35 36 37 39 41 42 44 46 47 48 49
    \item Validation subjects: 8 19 21 38 40
    \item Test subjects: 2 6 9 10 15 22 23 28 43 45
\end{itemize}

\begin{figure*}[htb]
    \centering
    \includegraphics[width=0.8\linewidth]{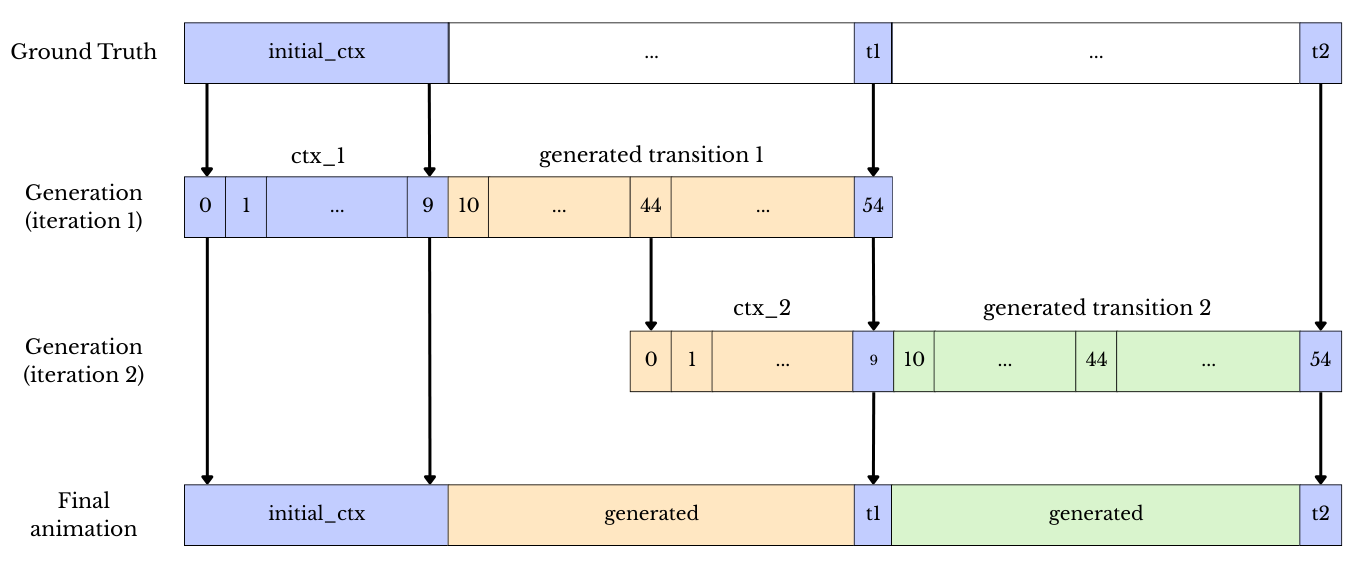}
    \caption{Diagram showing two iterations of the autoregressive idle motion generation process. t1 and t2 refer to the target frames of transition 1 and 2, respectively}
    \label{fig:gen_process}
\end{figure*}

\section{Towards standardised benchmarks}\label{sec:data_validation}
Together with publicly releasing StayStill, we also propose a evaluation pipeline based both on numerical and user-based metrics. We believe that having a standardised evaluation pipeline can be of great help when developing and comparing idle animation generation techniques. Firstly, for the numerical evaluation, the data was used to train a state-of-the-art transformer for the motion in-betweening task, but specifically centred around idle motion, and its results were numerically compared to other baselines.

Secondly, an idle animation generator was additionally constructed using the same transformer as a base. Then, the output of the neural network was rendered, and analysed in an extensive user study, to measure the perceptual naturalness of the generated motion, comparing it to the motion generated by other baselines. 

It is important to note that the objective of this paper is not to search for the best possible architectures and parameters for an idle animation generator. Instead, we search a viable first technique based on deep learning that generates idle animations to compare it to other baselines.
The code, data and the models used for both evaluations are publicly available alongside the dataset.

\subsection{Experimental details}\label{sec:experimental_details}
For both the user-based and the numerical comparisons, we used retrained versions of the two-stage transformer network by Qin et al. \cite{qin2022motion}, which is formed by two separate transformer networks. The first one, named the context transformer, is trained in a sequence-to-sequence manner, using motion sequences from the dataset as both input and target output. The second transformer uses the output of the first transformer as input and the original dataset sequences as target output, performing a refinement stage. Theoretically, this permits each transformer to separately learn and perform two different tasks: the context transformer models the coarse motion dynamics and the detail transformer learns to model the details of the motion.

To train the networks that we used in the evaluations, we always used the proposed standardised dataset splits, and trained each network over 200 epochs, using a batch-size of 32 and the same learning rate parameters as in the original paper (0.0625 and 0.025 for the context and the detail model, respectively, with 8000 iterations for warmup). Note that all models were trained using sequences of up to 30 frames, as training with maximum 45 frames did not yield better results. The model operates at the same frame-rate as the data in StayStill, which is 30 frames per second, and it uses 6D angle representations internally.

Moreover, for the user-based study, since we needed 10 seconds long animations to show to the participants, the generation process was conducted in an autoregressive manner, as shown in Figure \ref{fig:gen_process}. The generator was provided with 10 initial ground-truth context frames and a target frame located 45 frames ahead; once the in-between transition was produced, the last 10 frames of that segment serve as the context for the next iteration. In practice, the system is conditioned on the first 10 ground-truth frames and subsequently on one additional ground-truth keyframe every 45 frames, while all intermediate frames are generated through infilling.

During training, the data preprocessing starts by centring each initial context at the world origin before feeding it to the network, and rotating it to face the forward direction. This means that, during inference, each transition must be generated starting from the world origin and facing forward. Because of this, the final individual generated parts need to be added an offset and rotated to create a continuous animation.

Moreover, we found that the first frame generated by the neural network sometimes performs a jump that creates a discontinuity that is very visible. To avoid this, we applied a post-process phase to correct these discontinuities while trying to maintain the original dynamics of the motion with the least amount of modification possible.

\begin{figure*}[htb]
    \centering
    \includegraphics[width=0.8\linewidth]{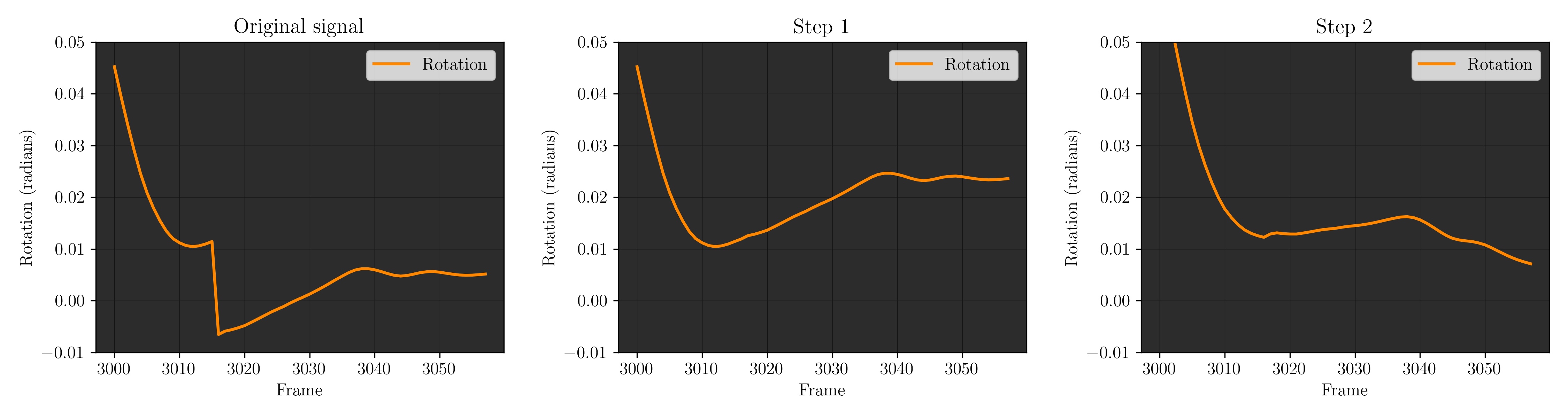}
    \caption{The two-step signal post-processing technique}
    \label{fig:correction_process}
\end{figure*}

The post-process is performed as follows: for each joint rotation, we first shift the entire signal to match the tendency of the previous one. This way, the shape of the original signal is not altered, but just shifted to remove the jump. However, this may generate error accumulation on the long run, since the shifted rotations might take undesirable values and, therefore, physically implausible or unnatural poses. Consequently, in a second phase, we add the inverse effect of the initial shift, equally divided between all frames of the transition. For example, if a rotation was initially shifted up 5 degrees, the consequent frames will shift down the rotation $5/45 = 0.11$ degrees in each frame, so the signal ends in the same position as the original one. The final animation preserves the general dynamics of the original signal, and does not accumulate error, as it exactly matches the original keyframes. Figure \ref{fig:correction_process} shows an example of the correction algorithm.

\begin{figure}[htb]
    \centering
    \includegraphics[width=0.66\linewidth]{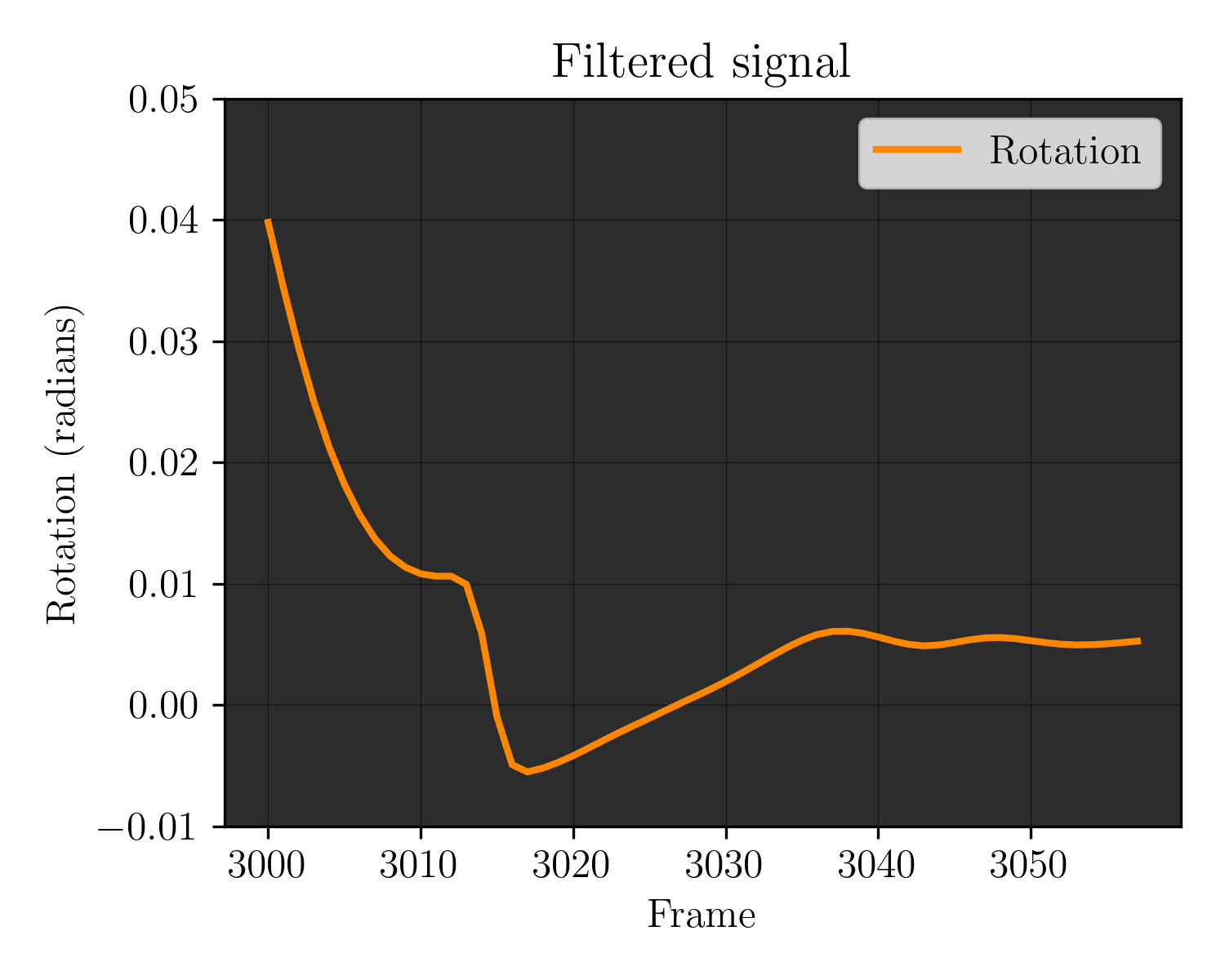}
    \caption{If the signal is filtered with a gaussian kernel, the resulting curve does not maintain the original motion dynamics}
    \label{fig:gaussian_correction}
\end{figure}

Although this post-processing method modifies the signal, it preserves the original signal properties much better than standard smoothing algorithms. Such algorithms often distort the dynamics of the signal, potentially changing the perception of the animations. For instance, for the original signal provided in Figure \ref{fig:correction_process}, a kernel size 7 Gaussian smoothing algorithm would create a ``U-shaped'' signal, as shown in Figure \ref{fig:gaussian_correction}, while our process maintains the original ascending behaviour of the original signal. However, we believe that a holistic motion generation method should reduce the generated jumps, and keep post-processing to a bare minimum. To support reproducibility, we release the pretrained models and generation code alongside the data. 

\subsection{Numerical evaluation} \label{sec:numerical_evaluation}
As aforementioned, for the numerical evaluation, we used the transformer with no post processing or autoregression applied. We then measured its performance in different time windows (5, 15, 30 and 45 frames) against standard baseline techniques as spherical linear interpolation (SLERP), spherical linear interpolation with quadratic ease-in-out (SLERP-Q) and 0 velocity (0-vel).

We trained the models in each part of our idle animation dataset. In other words, we trained and tested each model using just the idle actions instances, general idle instances or idle with a phone instances, separately. For each subset, we employed a 5-fold cross validation scheme in order to reduce bias, by separating a different test-set for each fold.

The train, validation and test sets were done subject-wise, meaning that the frames that belong to one subject were contained in just one of the train, validation or test sets, avoiding cross-contamination. In each fold, we randomly separated 35 subjects for the train partition, 5 for validation and 10 for testing (70\% train, 10\% validation, 20\% test). Each specific split that was used can be found in Appendix \ref{app:splits}. Note that the first split is the standardised split that we proposed in Section \ref{sec:officialSplit}.

We evaluated the models and the baselines using 3 widely used metrics: L2 loss on joint positions shown in meters (L2P), quaternion L2 loss with sign-correction (L2Q) and Normalised Power Spectrum Similarity (NPSS) \cite{gopalakrishnan2019neural}. NPSS is a metric that complements MSE or L2 loss that evaluates the long-term predicting ability of a model. We used the evaluation code provided by Qin et al. \cite{qin2022motion} and we extended it to work with SLERP-Q; we also provide this modified numerical evaluation code.

Tables \ref{tb:comparison_idle}, \ref{tb:comparison_phone} and \ref{tb:comparison_actions} show the numerical results obtained by the baselines and the transformer neural network, alongside the standard deviations. Since the three tables show similar results but for different data subsets, tables referring to the idle with a phone and idle action subsets (tables \ref{tb:comparison_phone} and \ref{tb:comparison_actions}) can be found in Appendix \ref{app:numerical_comparison} for improved readability.

\begin{table*}[htb]
\centering
\footnotesize
\setlength{\tabcolsep}{2pt}
\begin{tabular}{lcccccccccccc}
\hline
\textbf{General idle}      & \multicolumn{4}{c}{\textbf{L2P}}                                                                                                                                      & \multicolumn{4}{c}{\textbf{L2Q}}                                                                                                                                    & \multicolumn{4}{c}{\textbf{NPSS}}                                                                                                                              \\ \hline
\textbf{Transition length} & \multicolumn{1}{c}{\textbf{5}}            & \multicolumn{1}{c}{\textbf{15}}           & \multicolumn{1}{c}{\textbf{30}}          & \multicolumn{1}{c}{\textbf{45}} & \multicolumn{1}{c}{\textbf{5}}         & \multicolumn{1}{c}{\textbf{15}}           & \multicolumn{1}{c}{\textbf{30}}           & \multicolumn{1}{c}{\textbf{45}} & \multicolumn{1}{c}{\textbf{5}}          & \multicolumn{1}{c}{\textbf{15}}        & \multicolumn{1}{c}{\textbf{30}}        & \multicolumn{1}{c}{\textbf{45}} \\ \hline
0-velocity                 & \multicolumn{1}{l}{.900 \tiny{$\pm.15$}}          & \multicolumn{1}{l}{2.206 \tiny{$\pm.4$}}           & \multicolumn{1}{l}{3.826 \tiny{$\pm.73$}}         & 5.074 \tiny{$\pm.98$}                     & \multicolumn{1}{l}{.066 \tiny{$\pm.01$}}       & \multicolumn{1}{l}{.150 \tiny{$\pm.02$}}          & \multicolumn{1}{l}{.246 \tiny{$\pm.03$}}          & .316 \tiny{$\pm.04$}                     & \multicolumn{1}{l}{.0002 \tiny{$\pm0$}}          & \multicolumn{1}{l}{.002 \tiny{$\pm0$}}          & \multicolumn{1}{l}{.008 \tiny{$\pm0$}}          & .020 \tiny{$\pm0$}                        \\ \hline
SLERP                      & \multicolumn{1}{l}{\textbf{.197 \tiny{$\pm.03$}}} & \multicolumn{1}{l}{\textbf{.689 \tiny{$\pm.12$}}} & \multicolumn{1}{l}{1.605 \tiny{$\pm.28$}}         & 2.558 \tiny{$\pm.47$}                     & \multicolumn{1}{l}{\textbf{.022 \tiny{$\pm0$}}} & \multicolumn{1}{l}{\textbf{.066 \tiny{$\pm.01$}}} & \multicolumn{1}{l}{.128 \tiny{$\pm.02$}}          & .180 \tiny{$\pm.03$}                     & \multicolumn{1}{l}{\textbf{.0001 \tiny{$\pm0$}}} & \multicolumn{1}{l}{\textbf{.001 \tiny{$\pm0$}}} & \multicolumn{1}{l}{\textbf{.006 \tiny{$\pm0$}}} & \textbf{.018 \tiny{$\pm0$}}               \\ \hline
SLERP-Q                    & \multicolumn{1}{l}{.259 \tiny{$\pm.04$}}          & \multicolumn{1}{l}{.763 \tiny{$\pm.13$}}          & \multicolumn{1}{l}{1.631 \tiny{$\pm.29$}}         & 2.530 \tiny{$\pm.46$}                     & \multicolumn{1}{l}{.024 \tiny{$\pm.01$}}       & \multicolumn{1}{l}{.068 \tiny{$\pm.01$}}          & \multicolumn{1}{l}{\textbf{.126 \tiny{$\pm.02$}}} & \textbf{.176 \tiny{$\pm.03$}}            & \multicolumn{1}{l}{\textbf{.0001\tiny{$\pm0$}}}  & \multicolumn{1}{l}{\textbf{.001 \tiny{$\pm0$}}} & \multicolumn{1}{l}{\textbf{.006 \tiny{$\pm0$}}} & \textbf{.018 \tiny{$\pm0$}}               \\ \hline
Neural Network             & \multicolumn{1}{l}{.609 \tiny{$\pm.37$}}          & \multicolumn{1}{l}{.885 \tiny{$\pm.37$}}          & \multicolumn{1}{l}{\textbf{1.525 \tiny{$\pm.4$}}} & \textbf{2.352 \tiny{$\pm.45$}}            & \multicolumn{1}{l}{.075 \tiny{$\pm.06$}}       & \multicolumn{1}{l}{.102\tiny{$\pm.06$}}           & \multicolumn{1}{l}{.154 \tiny{$\pm.06$}}          & .207 \tiny{$\pm.06$}                     & \multicolumn{1}{l}{\textbf{.0001 \tiny{$\pm0$}}} & \multicolumn{1}{l}{\textbf{.001 \tiny{$\pm0$}}} & \multicolumn{1}{l}{.007 \tiny{$\pm0$}}          & .020 \tiny{$\pm.01$}                     \\ \hline
\end{tabular}
\caption{Comparison of the transformer network and baselines on the general idle part}
\label{tb:comparison_idle}
\end{table*}

In short transitions of 5 frames, SLERP remains unbeaten in all metrics. Both in L2P and L2Q error, and in NPSS, the spherical linear interpolation always obtains the smallest values. However, when moving towards longer transitions of 30 and 45 frames, the L2P error of the transformer increases slower than that of SLERP, both in general idle and specially in idle actions. When it comes to L2Q error, SLERP-Q obtains the best results in general idle. Nonetheless, in the idle actions part, the transformer obtains better values, and on the other two parts, the results remain very similar. The SLERP-Q baseline obtains similar values as SLERP overall, and even obtains smaller error in idle actions. However, in some cases, like using a phone, it yields bigger errors. Finally, 0-velocity obtains the worst results in nearly all sections, and it especially worsens when the transitions are long, as expected. In summary, while SLERP remains the superior baseline for near-instantaneous transitions, the neural network demonstrates greater scalability and precision as temporal distance increases, particularly within the complex dynamics of specific idle actions.

\subsection{User-based evaluation}
In many subfields of character animation, the validity of using numerical metrics as universal measuring tools is widely questioned \cite{Crnek04032026, kucherenko2024evaluating, kucherenko2025evaluating, zhao2025ppmotion}. Capturing the nuances of human perception with these kinds of measures is an extremely complex problem, which has resulted in the use of human assessment methods to better evaluate animations.

One of the most prominent examples of the need of human perceptual evaluation is the GENEA challenge \cite{kucherenko2023genea}. GENEA has been an ongoing  challenge on co-speech gesture generation that has made a huge effort to establish a common evaluation pipeline based on human perception. Findings from the GENEA Challenge 2022 highlight a significant divergence between numerical metrics and human perception \cite{kucherenko2024evaluating}.

Taking into account that idle animation generation is an underdeveloped research field, and no thorough analysis has been carried out to assess the usability of numerical metrics in this area, we also carried out an extensive user study. This way, our objective is to assess whether the animations generated by a deep learning system trained on StayStill are perceptually more realistic than other baselines that achieve strong scores in numerical metrics that have been presented in Section \ref{sec:numerical_evaluation}.

\subsubsection{The user study}
We conducted a user study consisting of pairwise comparisons of different motion clips. This has two main objectives: firstly, it defines an initial benchmarking process that can be adopted and extended by future work. Secondly, it evaluates the performance of a deep learning system trained with the data by measuring how people perceive the animations generated by the neural network. Therefore, we compared the following systems between them:

\begin{itemize}
	\item Ground Truth (GT): We used this condition as an upper bound for perceptual quality. It contains the original sequences from the dataset.
	\item Linear Interpolator (SLERP): We used linearly interpolated transitions between keyframes, which produce strong numeric results as a baseline, to compare to it in perceptual terms.
	\item SLERP with a quadratic in-out easing (SLERP-Q): We also compared the slightly more complex baseline that softens the edges of the linearly interpolated signal to obtain smoother movement.
	\item The two-stage generator (NN): The results produced by the neural network defined in Section \ref{sec:experimental_details}, with the autoregression and post processing enabled.
	\item Ground Truth with added Gaussian Noise (NOISE): We used this method as a lower bound for perceptual quality.
\end{itemize}

The study consisted of the following: each participant was shown 20 pairs of videos, and for each pair, they had to indicate their preference. More specifically, they were asked to respond to the following question: ``Which video is the most natural and realistic one?''. Then, the user selected a response out of 5 possible options: \textit{video 1, clear preference; video 1, slight preference; no preference; video 2, slight preference; video 2, clear preference}.

We employed Blender \cite{blender} to render the videos, using a common 3D humanoid model (the Y-bot from the Mixamo \cite{mixamo} repository). We removed the fingers of the model, since the original animations did not provide high-quality finger data. For each system that we were evaluating, we rendered the test set of the standardised dataset split. The rendered videos were 10 seconds long, and the clips were evenly spaced, so the entire test set was uniformly covered.

Then, we filtered out the inconsistent animations based on the ground truth: if the arms passed through the body of the avatar, all five corresponding videos for that motion sequence were discarded.

Following the evaluation practices of the aforementioned GENEA challenge \cite{kucherenko2023genea}, some final animations were discarded from the end of the test set, so the final pool of videos contained 50 instances for each system, resulting in 250 samples in total. During the study, each participant was shown 20 random pairs of videos taken from the pool. This ensured coverage of the entire test set, while avoiding excessive participant fatigue.

To ensure a fair comparison, all baselines in this study were evaluated against the same ground-truth clips using identical keyframes. While this deterministic approach was necessary for our current benchmarks, it may require adjustment as the field moves toward stochastic (probabilistic) generation. For such models, comparison remains feasible by either randomly pairing their outputs or by pairing the outputs that use the same control signals extracted from the StayStill test set.

The user study was crafted in three different languages: Basque, Spanish and English, so each participant could choose the language they were most comfortable with. Moreover, the questionnaire contained two attention checks in random positions, one at the beginning and another one in the end, in which users were asked to select a specific option. If a user failed one of the attention checks, all of their responses were discarded.

\begin{figure}[htb]
    \centering
    \includegraphics[width=\linewidth]{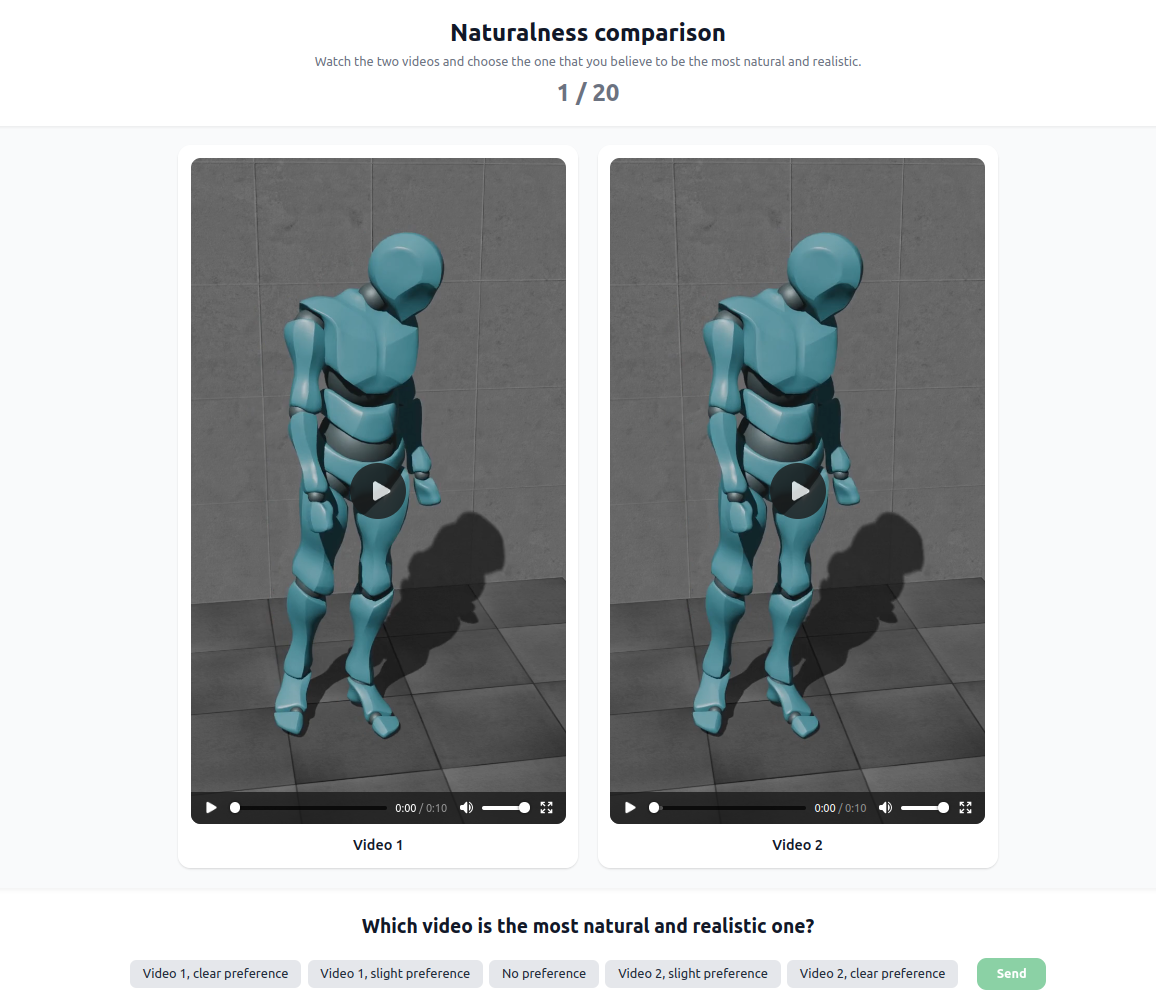}
    \caption{Image of the user study GUI and videos}
    \label{fig:user_study}
\end{figure}

The first two comparisons of the questionnaire were also fixed for calibration, serving to familiarise participants with the exercise: they showed a very obviously noisy animation against a ground truth clip, and a linearly interpolated clip against ground truth. This way, we prepared the participants with two easy exercises before starting the real test. The results of this two questions were not taken into account in the final statistics. Figure \ref{fig:user_study} shows a capture of the user interface of the study.

This setting enables to statistically analyse the results using a Bradley-Terry model \cite{bradley1952rank}. The evaluation process is based on the robust user-based benchmarking method present in the GENEA Challenges \cite{kucherenko2023genea}, and, more recently, the GENEA Leaderboard evaluation system \cite{nagy2026towards} and other gesture generation papers, such as \cite{alexanderson2023listen}.

\subsubsection{Participant recruitment}
We recruited participants among university students and teachers. Participation was entirely voluntary. After removing incomplete survey responses and excluding participants that failed any of the attention checks, we ended up with a sample of 118 participants. The demographic characteristics are summarised as follows:
\begin{itemize}
	\item Gender: 75.39\% male, 23.69\% female, 0.92\% non-binary
	\item Test language: 65.55\% Basque, 28.67\% Spanish, 5.78\% English
	\item Age: mean: 24.94, standard-deviation: 8.47, median 22
\end{itemize}

\subsubsection{Results}
The user study contained 1971 of evaluated pairs. Specifically, the appearance counts of each system are as follows: GT - 791, SLERP - 765, SLERP-Q - 782, NN - 771, NOISE - 833. The exact user preference confusion matrices can be found in Appendix \ref{app:user_preferences}. The results were analysed by estimating Elo scores using maximum likelihood estimation, following the same method as in the GENEA Leaderboard \cite{nagy2026towards}. A Bradley-Terry model was fitted using logistic regression on pairwise comparisons, by giving different weights to ties ($weight = 1$), weak wins —or the slight preference option— ($weight = 2$) and strong wins —or the clear preference option— ($weight = 4$). A fixed Elo scale was used with base 10, scale 400 and initial Elo score 1000. Consequently, the probability of a method A to beat method B is calculated as follows:

\begin{equation}
P(A beats B) = \frac{1}{1+10^{-(Elo_A - Elo_B)/400}}
\end{equation}

We estimated uncertainty in the Elo ratings by bootstrapping at the user level for 10,000 iterations. This involved sampling users with replacement and incorporating all their corresponding pairwise comparisons into each iteration.

Table \ref{tb:elo_scores} shows the Elo scores and win-rates (W/R) and their respective standard deviations based on the user-level bootstrapping. The table shows ground truth as the system with the highest Elo score, and the noise with the lowest, as expected. Then, the output of the neural network and SLERP-Q rank in second place, followed by SLERP. This means that, even though SLERP has numerically strong results, it falls back when it comes to human perception, meaning that the animations generated by the neural network have better perceptual quality. It is noteworthy that SLERP-Q obtains much better results than SLERP, as it creates visually more pleasing transitions, by removing the linearity of the generated animations. Figure \ref{fig:elo_forest} (Appendix \ref{app:elo_forest}) graphically shows the Elo results and the standard deviations.

\begin{table}[htb]
\centering
\begin{tabular}{|l|cc|cc|}
\hline
\textbf{System}  & \textbf{Elo} & \textbf{Std Elo} & \textbf{W/R} & \textbf{Std W/R} \\ \hline
\textbf{GT}      & 1179.36      & 15.71            & 77.74\%      & 1.81\%           \\
\textbf{NN}      & 1083.73      & 11.30            & 65.35\%      & 1.68\%           \\
\textbf{SLERP-Q} & 1083.70      & 13.50            & 64.84\%      & 2.03\%           \\
\textbf{SLERP}   & 913.38       & 17.47            & 38.15\%      & 2.73\%           \\
\textbf{NOISE}   & 739.82       & 24.25            & 14.27\%      & 2.13\%           \\ \hline
\end{tabular}
\caption{Elo scores and win-rates of the 5 proposed methods based on the user responses on pairwise comparisons, alongside the standard deviations calculated with 10000 bootstrap iterations}
\label{tb:elo_scores}
\end{table}
For pairwise comparisons, we computed 95\% confidence intervals for the differences in Elo ratings between systems using the empirical bootstrap distribution of their rating differences. For each system pair, the interval was obtained from the 2.5th and 97.5th percentiles of the bootstrap differences; a comparison was considered statistically significant if this interval did not include zero, indicating that the observed performance gap was unlikely to be due to sampling variability. Using this test, all differences between models were statistically significant at the 95\% level, except for the difference between the neural network and SLERP-Q, whose confidence interval included zero. For all other system pairs, no bootstrap samples crossed zero in 10,000 replicates, implying $p < 1\times10^{-4}$ and demonstrating the robustness of the observed ranking differences.

\section{Conclusions and future work}\label{sec:conclusions}
Idle animation generation using deep learning is an under explored research topic. With this work we take a step towards enabling research on idle animation synthesis using deep learning by providing a dataset for training and a methodology for evaluating the models.

First, we present and openly publish StayStill, a large-scale 3D idle animation dataset prepared for training deep learning neural networks. It includes 6 hours of idle motion data, divided into three main groups: regular idle sequences, idle sequences while using a phone, and idle actions. The idle actions subset contains 18 classes, each with individually labelled action sequences for each type of action. The dataset contains recordings of 50 different subjects, which result in high variability and very different styles of idling and performing actions. This variability is highly beneficial, particularly given the specialized nature of a dataset centered on a specific scenario like idling. We believe that StayStill is a significant leap in enabling deep learning-based idle animation generation, as it is the first dataset of its kind, to the best of our knowledge.

Furthermore, drawing on motion in-betweening methodologies, we propose a dual evaluation pipeline for idle animation generators. It combines both a numerical and a user-based comparison, as numerical evaluations in animation rarely capture the full picture. Widely used numerical metrics, such as positional and rotational error, fail to correlate with user responses in our experiments. For instance, while spherical linear interpolation produces strong numerical results, comparable to the neural network and to SLERP-Q, it performs poorly in terms of perceptual realism. This discrepancy confirms that user-based evaluation remains the gold standard for final benchmarking.

Alongside openly and freely publishing all the annotated data in StayStill, we also open source the code and the pre-trained model of the proposed animation generator, which is based on creating realistic transitions between ground truth poses autoregressively and concatenating the generated parts into a long animation sequence. The code for the user-based evaluation is also openly available in the repository, as well as the responses of the users.

It is noteworthy that the animations do not contain finger data, due to the difficulties of providing high quality detections with markerless motion capture systems. While careful manual inspection has been conducted, the used setting also may introduce small artifacts like short jitter, foot sliding or self-penetrations when crossing arms, for example.

Moreover, the animation samples compared in the numerical and user evaluations were coupled to a common ground truth, which is not ideal if the generation has to be probabilistic. We believe that the user evaluation pipeline and code are nevertheless usable if the field gravitates towards probabilistic generation: the pairwise comparisons would need to be randomly done, and conditional generation would need to be conditioned on the comparable signals, calculated from the test set. In that case, the numerical metrics would also need to better compare the differences between distributions, instead of measuring the differences of specific clips, as MSE does. 

As future work, on the one hand, we plan to investigate the use other metrics to compare probabilistic generation that are more suitable to compare distributions, such as Frechet Motion Distance \cite{maiorca2022evaluating}. On the other hand, we plan to train an idle animation generator using StayStill, not based on motion in-betweening but by directly generating idle animations.

Exploring idle motion dynamics is a good way of adding realism and personality to virtual characters, by generating different ways and styles to perform idle movements. The person-wise and class-wise variability of StayStill will permit to develop rich systems that are able to produce stylistically diverse animations and characters.

\section{Acknowledgements}\label{sec:acknowledgements}
This work has been partially funded by the Basque Government, Spain, under Research Teams Grant number IT1812-26; the Spanish Ministry of Science (MCIU), the State Research Agency (AEI), the European Regional Development Fund (FEDER), under Grant number PID2021-122402OB-C21 (MCIU/AEI/FEDER, UE); and the University of the Basque Country (UPV/EHU) under grant PIF 23/07.


\printbibliography                

\newpage

\appendix
\section{Dataset detail}\label{app:motionData}
\begin{table}[htb]
\centering
\begin{tabular}{|l|l|l|l|}
\hline
{\textbf{Motion type}}             & {\textbf{Frames}} & {\textbf{Duration}} & {\textbf{Clips}} \\ \hline
Action: look up/sky               & 27,943                           & 15:31                              & 98                             \\ \hline
Action: look around               & 20,338                           & 11:17                              & 92                             \\ \hline
Action: look down/floor           & 17,817                           & 9:53                               & 85                             \\ \hline
Action: look shoes                & 15,623                           & 8:40                               & 75                             \\ \hline
Action: check watch               & 11,332                           & 6:17                               & 92                             \\ \hline
Action: check phone               & 21,237                           & 11:47                              & 89                             \\ \hline
Action: scratch head              & 13,199                           & 7:19                               & 92                             \\ \hline
Action: scratch arm               & 13,750                           & 7:38                               & 93                             \\ \hline
Action: scratch leg               & 11,510                           & 6:23                               & 77                             \\ \hline
Action: scratch back              & 10,900                           & 6:03                               & 66                             \\ \hline
Action: touch face/chin           & 14,852                           & 8:15                               & 93                             \\ \hline
Action: stretch arms              & 14,473                           & 8:02                               & 80                             \\ \hline
Action: stretch back              & 12,300                           & 6:50                               & 71                             \\ \hline
Action: rub eyes                  & 15,430                           & 8:34                               & 97                             \\ \hline
Action: yawn                      & 13,766                           & 7:38                               & 95                             \\ \hline
Action: look back (left)   & 10,253                           & 5:41                               & 69                             \\ \hline
Action: look back (right)  & 10,571                           & 5:52                               & 75                             \\ \hline
Action: balance l/r & 11,100                           & 6:10                               & 48                             \\ \hline
Action: balance r/l & 9,357                            & 5:11                               & 47                             \\ \hline
\textbf{Idle actions total}            & \textbf{275,751}                 & \textbf{2:33:11}                   & \textbf{1534}                  \\ \hline
\textbf{General Idle}                  & \textbf{181,846}                 & \textbf{1:41:01}                   & \textbf{50}                    \\ \hline
\textbf{Idle with a phone}             & \textbf{187,741}                 & \textbf{1:44:18}                   & \textbf{50}                    \\ \hline
\end{tabular}
\caption{Dataset details: recorded motion types, durations and clip quantities}
\label{tb:motionData}
\end{table}
\section{Train-val-test splits} \label{app:splits}
\begin{table}[htb]
\begin{tabular}{|l|p{5cm}|l|l|}
\hline
Split & Train                                                                                        & Validation  & Test                  \\ \hline
1        & 0 1 3 4 5 7 11 12 13 14 16 17 18 20 24 25 26 27 29 30 31 32 33 34 35 36 37 39 41 42 44 46 47 48 49  & 8 19 21 38 40      & 2 6 9 10 15 22 23 28 43 45 \\ \hline
2        & 0 1 2 3 4 5 6 9 10 11 13 15 16 17 18 19 20 21 23 24 25 26 28 29 30 32 34 35 38 39 41 42 44 40 45     & 22 33 37 43 49      & 7 8 12 14 27 31 36 46 47 48 \\ \hline
3        & 0 1 2 6 7 8 9 10 12 14 15 17 18 19 20 22 23 24 25 26 27 28 31 33 35 37 39 41 42 43 44 45 46 47 48   & 3 11 34 36 38      & 4 5 13 16 21 29 30 32 49 50 \\ \hline
4        & 2 4 5 6 8 9 10 11 13 14 15 18 19 20 21 22 23 24 27 29 30 31 32 33 35 36 40 42 43 44 45 46 47 48 49   & 7 12 16 28 39      & 0 1 3 17 25 26 34 37 38 41 \\ \hline
5        & 0 1 3 4 6 7 8 9 12 13 14 15 16 17 21 22 23 25 26 27 29 30 31 34 36 37 38 40 41 43 45 46 47 48 49    & 2 5 10 28 32      & 11 18 19 20 24 33 35 39 42 44 \\ \hline
\end{tabular}
\caption{The subjects contained in each of the train, validation and test splits in each fold.}
\end{table}
\clearpage
\section{User responses confusion matrices}\label{app:user_preferences}
\begin{table}[H]
\begin{tabular}{|l|l|l|l|l|l|}
 \hline
       & GT & NN & SLERP-Q & SLERP & NOISE \\ \hline
GT     & 0  & 47 & 51     & 46    & 26    \\ \hline
NN     & 39 & 0  & 29     & 58    & 29    \\ \hline
SLERP-Q & 36 & 54 & 0      & 61    & 29    \\ \hline
SLERP  & 19 & 24 & 14     & 0     & 24    \\ \hline
NOISE  & 7  & 23 & 25     & 24    & 0    \\  \hline
\end{tabular}
\caption{Confusion matrix showing weak wins in the user responses}
\end{table}

\begin{table}[H]
\begin{tabular}{|l|l|l|l|l|l|}
 \hline
       & GT & NN & SLERP-Q & SLERP & NOISE \\ \hline
GT     & 0  & 36 & 45     & 93    & 158   \\ \hline
NN     & 10 & 0  & 20     & 49    & 148   \\ \hline
SLERP-Q & 12 & 15 & 0      & 34    & 142   \\ \hline
SLERP  & 16 & 6  & 1      & 0     & 105   \\ \hline
NOISE  & 5  & 10 & 10     & 28    & 0    \\ \hline
\end{tabular}
\caption{Confusion matrix showing strong wins in the user responses}
\end{table}

\begin{table}[H]
\begin{tabular}{|l|l|l|l|l|l|}
 \hline
       & GT & NN & SLERP-Q & SLERP & NOISE \\ \hline
GT     & 0  & 58 & 46     & 29    & 12    \\ \hline
NN     & 58 & 0  & 67     & 42    & 7     \\ \hline
SLERP-Q & 46 & 67 & 0      & 81    & 10    \\ \hline
SLERP  & 29 & 42 & 81     & 0     & 11    \\ \hline
NOISE  & 12 & 7  & 10     & 11    & 0    \\ \hline
\end{tabular}
\caption{Confusion matrix showing ties in the user responses}
\end{table}

\clearpage
\section{Numerical results in the idle with a phone and idle actions subsets}\label{app:numerical_comparison}

\begin{table}[htb]
\footnotesize
\setlength{\tabcolsep}{2pt}
\begin{tabular}{lcccccccccccc}
\hline
\textbf{Idle with a phone} & \multicolumn{4}{c}{\textbf{L2P}}                                                                                                                                 & \multicolumn{4}{c}{\textbf{L2Q}}                                                                                                                               & \multicolumn{4}{c}{\textbf{NPSS}}                                                                                                                            \\ \hline
\textbf{Transition length} & \multicolumn{1}{c}{\textbf{5}}          & \multicolumn{1}{c}{\textbf{15}}         & \multicolumn{1}{c}{\textbf{30}}         & \multicolumn{1}{c}{\textbf{45}} & \multicolumn{1}{c}{\textbf{5}}        & \multicolumn{1}{c}{\textbf{15}}         & \multicolumn{1}{c}{\textbf{30}}         & \multicolumn{1}{c}{\textbf{45}} & \multicolumn{1}{c}{\textbf{5}}         & \multicolumn{1}{c}{\textbf{15}}        & \multicolumn{1}{c}{\textbf{30}}       & \multicolumn{1}{c}{\textbf{45}} \\ \hline
0-velocity                 & \multicolumn{1}{l}{0.645\tiny{$\pm.04$}}          & \multicolumn{1}{l}{1.596\tiny{$\pm.12$}}          & \multicolumn{1}{l}{2.838\tiny{$\pm.22$}}          & 3.814\tiny{$\pm.33$}                       & \multicolumn{1}{l}{0.040\tiny{$\pm0$}}        & \multicolumn{1}{l}{0.084\tiny{$\pm.01$}}          & \multicolumn{1}{l}{0.134\tiny{$\pm.02$}}          & 0.170\tiny{$\pm.01$}                       & \multicolumn{1}{l}{0.0001\tiny{$\pm0$}}          & \multicolumn{1}{l}{0.000\tiny{$\pm0$}}           & \multicolumn{1}{l}{0.009\tiny{$\pm.01$}}          & 0.008\tiny{$\pm0$}                         \\ \hline
SLERP                      & \multicolumn{1}{l}{\textbf{0.143\tiny{$\pm.01$}}} & \multicolumn{1}{l}{\textbf{0.505\tiny{$\pm.02$}}} & \multicolumn{1}{l}{\textbf{1.151\tiny{$\pm.07$}}} & 1.847\tiny{$\pm.12$}                       & \multicolumn{1}{l}{\textbf{0.012\tiny{$\pm0$}}} & \multicolumn{1}{l}{\textbf{0.040\tiny{$\pm0$}}} & \multicolumn{1}{l}{0.072\tiny{$\pm0$}}          & \textbf{0.100\tiny{$\pm.01$}}              & \multicolumn{1}{l}{\textbf{0.0000\tiny{$\pm0$}}} & \multicolumn{1}{l}{\textbf{0.0004\tiny{$\pm0$}}} & \multicolumn{1}{l}{\textbf{0.002\tiny{$\pm0$}}} & \textbf{0.007\tiny{$\pm0$}}                \\ \hline
SLERP-Q                    & \multicolumn{1}{l}{0.667\tiny{$\pm0$}}          & \multicolumn{1}{l}{0.849\tiny{$\pm.02$}}          & \multicolumn{1}{l}{1.291\tiny{$\pm.08$}}          & 1.904\tiny{$\pm.12$}                       & \multicolumn{1}{l}{0.064\tiny{$\pm.01$}}          & \multicolumn{1}{l}{0.080\tiny{$\pm0$}}            & \multicolumn{1}{l}{0.107\tiny{$\pm0$}}          & 0.137\tiny{$\pm.01$}                       & \multicolumn{1}{l}{0.0001\tiny{$\pm0$}}          & \multicolumn{1}{l}{0.0006\tiny{$\pm0$}}          & \multicolumn{1}{l}{0.003\tiny{$\pm0$}}          & 0.009\tiny{$\pm0$}                         \\ \hline
Neural Network             & \multicolumn{1}{l}{0.187\tiny{$\pm.34$}}          & \multicolumn{1}{l}{0.559\tiny{$\pm.33$}}          & \multicolumn{1}{l}{1.168\tiny{$\pm.28$}}          & \textbf{1.813\tiny{$\pm.27$}}              & \multicolumn{1}{l}{0.014\tiny{$\pm.04$}}        & \multicolumn{1}{l}{\textbf{0.040\tiny{$\pm.04$}}} & \multicolumn{1}{l}{\textbf{0.070\tiny{$\pm.03$}}} & \textbf{0.100\tiny{$\pm.03$}}              & \multicolumn{1}{l}{\textbf{0.0000\tiny{$\pm0$}}} & \multicolumn{1}{l}{\textbf{0.0004\tiny{$\pm0$}}} & \multicolumn{1}{l}{\textbf{0.002\tiny{$\pm0$}}} & \textbf{0.007\tiny{$\pm0$}}                \\ \hline
\end{tabular}
\caption{Comparison of the transformer network and baselines on the idle with a phone part}
\label{tb:comparison_phone}
\end{table}

\begin{table}[htb]
\footnotesize
\setlength{\tabcolsep}{2pt}
\begin{tabular}{lcccccccccccc}
\hline
\textbf{Idle actions}      & \multicolumn{4}{c|}{\textbf{L2P}}                                                                                                                                 & \multicolumn{4}{c|}{\textbf{L2Q}}                                                                                                                               & \multicolumn{4}{c|}{\textbf{NPSS}}                                                                                                                            \\ \hline
\textbf{Transition length} & \multicolumn{1}{c|}{\textbf{5}}          & \multicolumn{1}{c|}{\textbf{15}}         & \multicolumn{1}{c|}{\textbf{30}}         & \multicolumn{1}{c|}{\textbf{45}} & \multicolumn{1}{c|}{\textbf{5}}        & \multicolumn{1}{c|}{\textbf{15}}         & \multicolumn{1}{c|}{\textbf{30}}         & \multicolumn{1}{c|}{\textbf{45}} & \multicolumn{1}{c|}{\textbf{5}}         & \multicolumn{1}{c|}{\textbf{15}}        & \multicolumn{1}{c|}{\textbf{30}}       & \multicolumn{1}{c|}{\textbf{45}} \\ \hline
0-velocity                 & \multicolumn{1}{l|}{0.802\tiny{$\pm.10$}}          & \multicolumn{1}{l|}{2.027\tiny{$\pm.25$}}          & \multicolumn{1}{l|}{3.689\tiny{$\pm.42$}}          & 5.053\tiny{$\pm.54$}                       & \multicolumn{1}{l|}{0.098\tiny{$\pm.01$}}        & \multicolumn{1}{l|}{0.258\tiny{$\pm.02$}}          & \multicolumn{1}{l|}{0.488\tiny{$\pm.02$}}          & 0.678\tiny{$\pm.03$}                       & \multicolumn{1}{l|}{0.0003\tiny{$\pm0$}}          & \multicolumn{1}{l|}{0.0042\tiny{$\pm0$}}          & \multicolumn{1}{l|}{0.025\tiny{$\pm0$}}          & 0.071\tiny{$\pm0$}                         \\ \hline
SLERP                      & \multicolumn{1}{l|}{\textbf{0.209\tiny{$\pm.01$}}} & \multicolumn{1}{l|}{0.806\tiny{$\pm.04$}}          & \multicolumn{1}{l|}{1.838\tiny{$\pm.15$}}          & 2.830\tiny{$\pm.28$}                       & \multicolumn{1}{l|}{\textbf{0.030\tiny{$\pm0$}}} & \multicolumn{1}{l|}{0.126\tiny{$\pm.01$}}          & \multicolumn{1}{l|}{0.260\tiny{$\pm.01$}}          & 0.378\tiny{$\pm.02$}                       & \multicolumn{1}{l|}{\textbf{0.0001\tiny{$\pm0$}}} & \multicolumn{1}{l|}{0.0030\tiny{$\pm0$}}          & \multicolumn{1}{l|}{0.020\tiny{$\pm0$}}          & 0.059\tiny{$\pm0$}                         \\ \hline
SLERP-Q                    & \multicolumn{1}{l|}{0.256\tiny{$\pm.02$}}          & \multicolumn{1}{l|}{0.837\tiny{$\pm.05$}}          & \multicolumn{1}{l|}{1.789\tiny{$\pm.16$}}          & 2.704\tiny{$\pm.29$}                       & \multicolumn{1}{l|}{0.038\tiny{$\pm0$}}          & \multicolumn{1}{l|}{0.128\tiny{$\pm0$}}            & \multicolumn{1}{l|}{0.248\tiny{$\pm.01$}}          & 0.358\tiny{$\pm.02$}                       & \multicolumn{1}{l|}{0.0002\tiny{$\pm0$}}          & \multicolumn{1}{l|}{0.0031\tiny{$\pm0$}}          & \multicolumn{1}{l|}{0.020\tiny{$\pm0$}}          & 0.058\tiny{$\pm0$}                         \\ \hline
Neural Network             & \multicolumn{1}{l|}{0.295\tiny{$\pm.07$}}          & \multicolumn{1}{l|}{\textbf{0.632\tiny{$\pm.07$}}} & \multicolumn{1}{l|}{\textbf{1.353\tiny{$\pm.13$}}} & \textbf{2.230\tiny{$\pm.20$}}              & \multicolumn{1}{l|}{0.046\tiny{$\pm.01$}}        & \multicolumn{1}{l|}{\textbf{0.100\tiny{$\pm.01$}}} & \multicolumn{1}{l|}{\textbf{0.208\tiny{$\pm.01$}}} & \textbf{0.319\tiny{$\pm.02$}}              & \multicolumn{1}{l|}{\textbf{0.0001\tiny{$\pm0$}}} & \multicolumn{1}{l|}{\textbf{0.0019\tiny{$\pm0$}}} & \multicolumn{1}{l|}{\textbf{0.016\tiny{$\pm0$}}} & \textbf{0.052\tiny{$\pm0$}}                \\ \hline
\end{tabular}
\caption{Comparison of the transformer network and baselines on the idle actions part}
\label{tb:comparison_actions}
\end{table}

\section{Graphical representation of Elo scores}\label{app:elo_forest}
\begin{figure}[htb]
    \centering
    \includegraphics[width=2.0\linewidth]{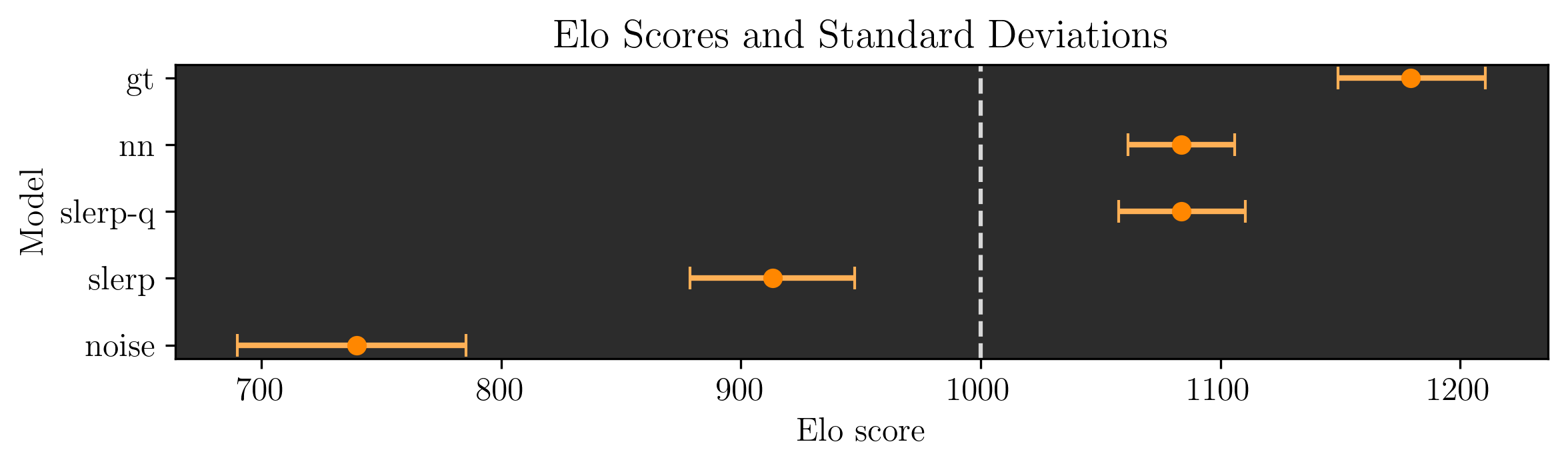}
    \caption{Graphical representation of the Elo scores and standard deviations calculated with 10000 bootstrap iterations.}
    \label{fig:elo_forest}
\end{figure}

\end{document}